 \documentclass[preprint,aps]{revtex4}

\usepackage{graphicx}
\usepackage{dcolumn}

\begin{document}

\preprint{Lebed-PRB}

\title{Ginzburg-Landau slopes of the anisotropic upper critical magnetic field 
and band parameters in the superconductor (TMTSF)$_2$ClO$_4$}

\author{A.G. Lebed}

\affiliation{Department of Physics, University of Arizona, 1118 E.
4-th Street, Tucson, AZ 85721, USA}
\affiliation{L.D. Landau Institute for Theoretical Physics,
2 Kosygina Street, 117334 Moscow, Russia}

\begin{abstract}
We theoretically determine the Ginzburg-Landau slopes of the anisotropic 
upper critical magnetic field in a quasi-one-dimensional superconductor and 
correct the previous works on this 
issue.
By using the experimentally measured values of the Ginzburg-Landau slopes 
in the superconductor (TMTSF)ClO$_4$, we determine band parameters of its 
electron spectrum.
Our main result is that the so-called quantum dimensional crossover has to 
happen in this material in magnetic fields, $H = 3-8 \ T$, which are much lower
than the previously 
assumed.
We discuss how this fact influences metallic and superconducting properties
of the (TMTSF)$_2$ClO$_4$.
 \\ \\ PACS numbers: 74.70.Kn
\end{abstract}


\maketitle

\pagebreak

Since a discovery of the field-induced spin-density-wave (FISDW) phase 
diagrams$^{1,2}$, high magnetic field properties of organic superconductors (TMTSF)$_2$X (X=ClO$_4$, PF$_6$, AsF$_6$, etc.) have been intensively 
studied$^{3,4}$. 
Phase transitions from metallic to the FISDW phases, exhibiting three-dimensional
quantum Hall effect, were successfully explained$^{5-11}$ in terms of the simplest 
quasi-classical $3D \rightarrow 2D$ dimensional 
crossover$^3$.
More complicated $3D \rightarrow 1D \rightarrow 2D$ quasi-classical dimensional
crossovers in a magnetic field successfully explain such phenomena in a metallic
phase as the Lebed Magic Angles (LMA)$^{12,13}$ and the Lee-Naughton-Lebed 
(LNL) oscillations$^{14,15}$.
The characteristic feature of the quasi-classical dimensional crossovers is that
a typical size of electron orbits in a magnetic field is much larger than inter-chain
and inter-plane distances in these layered quasi-one-dimensional (Q1D)
conductors.

Other dimensional crossovers - the quantum ones$^3$ - were suggested in 
Refs. 16-18 to demonstrate the Reentrant superconductivity phenomenon$^{16}$, 
where high magnetic fields can improve superconducting 
pairing.
Under condition of the quantum dimensional crossover, a typical size of electron
trajectories in a magnetic field becomes of the order or even less than interlayer
distance in (TMTSF)$_2$X conductors$^{16,19}$.
 Note that the quantum dimensional crossovers have been supposed to happen
 in magnetic fields of the order of $10-20 \ T$, parallel to conducting layers
 of  (TMTSF)$_2$X materials.
 
 The main goal of our Letter is to determine carefully band parameters of Q1D 
 electron spectrum of the conductor 
 (TMTSF)$_2$ClO$_4$ $^{20}$,
\begin{equation}
\epsilon({\bf p})= - 2t_a \cos(p_x a/2) - 2 t_b \cos(p_y b) - 2t_c
\cos (p_z c^*),
\end{equation}
where $t_a \gg t_b \gg t_c$ correspond to electron hoping integrals along
${\bf a}$ , ${\bf b}$, and ${\bf c^*}$ axes, respectively.
Using the determined band parameters, we show that the quantum dimensional 
crossover in the conductor (TMTSF)$_2$ClO$_4$ happens at much lower 
magnetic fields, 
$H \simeq 3-8 \ T$.
We discuss how this fact influence its magnetic properties in metallic and
superconducting phases and discuss the related experimental
data.
Below, we simplify electron spectrum (1) near two slightly corrugated
sheets of Q1D Fermi surface (FS) as
\begin{equation}
\delta \epsilon^{\pm}({\bf p})= \pm v_F(p_x \mp p_F) - 2t_b \cos(p_y b) 
- 2t_c \cos(p_z c^*) ,
\end{equation}
where +(-) stands for right (left) sheet of Q1D FS, $v_F=t_a a / \sqrt{2}$;
$\hbar \equiv 1$.

Let us consider electron motion in a magnetic field, perpendicular to
conducting chains and parallel to conducting layers,
\begin{equation}
{\bf H} = (0,H,0), \ \ {\bf A} = (0,0,-Hx),
\end{equation}
In accordance with Ref.16, electron spectrum (2) is "two-dimensionalized" 
in a magnetic field (3). 
More specifically, electrons are characterized by free unrestricted
motion within conducting $({\bf a},{\bf b})$ plane, whereas their motion along
${\bf z}$ axis is periodic and restricted$^{16}$:
\begin{equation}
z(t,H) = l_{\perp}(H) \ c^* \cos(\omega_c t) , \ \ l_{\perp}(H)=2t_c/\omega_c ,
\end{equation} 
where $\omega_c = ev_F Hc^*/c$.
By using quantum mechanical methods, it is possible to show$^{21,19,16}$ that 
the quantum $3D \rightarrow 2D$ dimensional crossover happens if a size of 
the quasi-classical orbit (4) is approximately in the range between $c^*$ and $c^*/2$ ,
\begin{equation}
l_{\perp}(H) \simeq 0.5 - 1.0.
\end{equation}
Classically, this corresponds to situation, where either electron orbits from two neighboring conducting layers do not intersect each other or do not intersect
neighboring layers, respectively.

Here, we express a value of the dimensionless parameter $l_{\perp}(H)$ in
terms of ratio of electron hoping integrals along ${\bf z}$ and ${\bf x}$ 
axes.
It is possible to show that 
\begin{equation}
l_{\perp}(H) = \frac{2 \sqrt{2}}{\pi} \frac{\phi_0}{a c^* H} \frac{t_c}{t_a} \simeq 
\frac{2 \times 10^3}{H(T)} \frac{t_c}{t_a}, 
\end{equation}
where $H(T)$ is a magnetic field, measured in Teslas.
Let us first use values of the parameters of the electron spectrum (1) of 
(TMTSF)$_2$ClO$_4$, accepted in literature$^{22}$, $t_a = 1200 \ K$
and $t_c = 7 \ K$.
In this case, as it follows from Eqs.(5),(6), the quantum $3D \rightarrow 2D$ 
dimensional crossover happens approximately at 
$H \geq H^* \simeq 12 - 23 \ T$.
In this Rapid Communication, we show that in reality $t_c \simeq 2 - 2.3 \ K$ 
and $t_a \simeq 1340 - 1130 \ K$, which result in the quantum dimensional crossover
at $H \geq H^* \simeq 3-8 \ T$.

Below, we derive the Ginzburg-Landau (GL) slope for the upper critical
magnetic field, parallel to {\bf b} axis of a singlet s-wave Q1D superconductor
with the electron spectrum (2).
For this purpose, we rewrite the so-called gap equation of Refs.16,23 in the
following way:
\begin{equation}
\Delta(x) = \frac{g}{2}  \int_{|z| > d} 
\frac{2 \pi T dz}{v_F \sinh \bigl( \frac{ 2 \pi T |z|}{v_F} \bigl)} 
J_0 \biggl[ \frac{2 t_c \omega_c}{v^2_F} z (z+2x) \biggl]  \Delta (x+z),
\end{equation}
where $g$ ia an effective electron coupling constant, $d$ is a cut-off
distance.
[Note that in Eq.(7) we disregard quantum effects of an electron motion in
a magnetic field in the extended Brillouin zone and, thus, replace the
functions $\sin[ \omega_c z/2v_F]$ and $\sin[ \omega_c (z+2x)/2v_F]$
by their arguments.
Here and everywhere below, we also disregards the Pauli paramagnetic 
destructive effects against superconductivity.]

The next step of derivation of the GL slop is to take into account that in 
the GL region, $(T_c -T)/T_c \ll 1$, $v_F /2 \pi T_c \ll v_F/ \sqrt{t_c \omega_c}$,
where $T_c$ is the superconducting transition temperature in the absence
of a magnetic field.
Therefore, we can expend the integral equation (7) with respect to a small 
parameter, $|z| \sim v_F / 2 \pi T_c$.
As a result of such expansion procedure, we obtain the following differential
equation:
\begin{eqnarray}
&&\biggl[ -\frac{d^2 \Delta(x)}{dx^2} + x^2 \frac{8 t^2_c \omega^2_c}{v^4_F}  
\Delta(x)   \biggl]      \int^{\infty}_0 
\frac{\pi  T_c z^2 dz}{v_F \sinh \bigl( \frac{2 \pi T_c z}{v_F} \bigl)}
\nonumber\\
&&+\biggl[ \frac{1}{g} -  \int^{\infty}_d  \frac{2 \pi T dz}{v_F \sinh \bigl( \frac{ 2 \pi T z}{v_F} \bigl)} \biggl] \Delta (x) = 0.
\end{eqnarray}
If we take into account that
\begin{equation}
 \frac{1}{g} =  \int^{\infty}_d  \frac{2 \pi T_c dz}{v_F \sinh \bigl( \frac{ 2 \pi T_c z}{v_F} \bigl)}  = 0 ,
\end{equation}
then we can rewrite Eq.(8) in the following way:
\begin{eqnarray}
&&- \xi^2_x \frac{d^2 \Delta(x)}{dx^2} + \biggl( \frac{2 \pi H}{\phi_0} \biggl)^2 \xi^2_z x^2 
\Delta(x) - \tau \Delta(x) = 0 ,
\nonumber\\
&&\xi^2_x = \frac{7 \zeta(3) v^2_F}{16 \pi^2 T^2_c}  , \  \xi^2_z = \frac{7 \zeta(3) t^2_c (c*)^2}{8 \pi^2 T^2_c} , \ \tau = \frac{T_c-T}{T_c},
\end{eqnarray}
where $\phi_0 = \pi \hbar c /e$ is the flux quantum, $\xi_x$ and $\xi_z$ are the coherence lengths along ${\bf a}$ and ${\bf c^* }$ axes,
correspondingly.
Note that above we use the following relationship:
\begin{equation}
\int^{\infty}_0 \frac{z^2 dz}{\sinh(z)} = \frac{7 \zeta(3)}{2},
\end{equation}
where $\zeta(n)$ is the Reimann zeta function$^{24}$.

To find the GL slope of the upper critical field along ${\bf b}$ axis, we need to
determine the lowest energy level of the Schrodinger-like GL equation (10).
After standard calculations, we obtain
\begin{equation}
H^b_{c2} = \frac{\phi_0}{2 \pi \xi_x \xi_z} \biggl( \frac{T_c-T}{T_c} \biggl) =
\frac{8 \pi^2 c \hbar T^2_c}{7 \zeta(3) e t_a t_c a c^*} \biggl( \frac{T_c-T}{T_c}
\biggl) .
\end{equation}
It is important that the GL slope of the upper critical field along ${\bf c}$ axis
for a singlet s-wave Q1D superconductor with electron spectrum (2) can be
obtained from Eq.(12) by using the following substitutions:
\begin{equation}
\xi_z \rightarrow \xi_y, \ t_c \rightarrow t_b, \ c^* \rightarrow b .
\end{equation}
As a result,
\begin{equation}
H^c_{c2} = \frac{\phi_0}{2 \pi \xi_x \xi_y} \biggl( \frac{T_c-T}{T_c} \biggl) =
\frac{8 \pi^2 c \hbar T^2_c}{7 \zeta(3) e t_a t_b a b} \biggl( \frac{T_c-T}{T_c}
\biggl) .
\end{equation}

Let us rewrite Eq.(13) of Ref.25, determining the upper critical field along
${\bf a}$ axis of a singlet d-wave Q1D superconductor, for s-wave case,
\begin{eqnarray}
\Delta(y) = &&\frac{g}{2} \biggl<  \int_{|z| > d} 
\frac{2 \pi T dz}{v_F \sinh \bigl( \frac{ 2 \pi T |z|}{v_F} \bigl)} 
\Delta \biggl[ y+\frac{v_y(p_y)}{v_F}z \biggl]
\nonumber\\
&&\times J_0 \biggl( \frac{2 t_c \omega_c}{v^2_F} z \biggl[ 2y 
+ \frac{v_y(p_y) z}{v_F} \biggl] \biggl) \biggl>_{p_y} ,
\end{eqnarray}
where $v_y(p_y) =  2 t_b b \sin(p_y b)$, $< ... >_{p_y}$ stands for averaging 
procedure over momentum component $p_y$.
By using the same method, as for determination of the GL slope for
${\bf H} \parallel {\bf b}$, we obtain the following GL slope for the upper
critical along ${\bf a}$ axis:
\begin{equation}
H^a_{c2} = \frac{\phi_0}{2 \pi \xi_y \xi_z} \biggl( \frac{T_c-T}{T_c} \biggl) =
\frac{4 \pi^2 c \hbar T^2_c}{7 \zeta(3) e t_b t_c b c^*} \biggl( \frac{T_c-T}{T_c}
\biggl) .
\end{equation}
 
 We stress that Eqs.(12),(14),(16) define the GL slopes of the upper
 critical fields in a singlet s-wave Q1D superconductor with the electron
 spectrum (1),(2) for all principal directions of a magnetic field. 
 These equations correct the previous results of Ref.26 and contain additional
 common factor 2/3 comparable to the corresponding equations 
 of Ref. 26.
 As it follows from general theory$^{27}$, for a singlet d-wave like Q1D
 superconductor (1),(2) with order parameter,
 \begin{equation}
 \Delta({\bf p}) = \sqrt{2} \Delta \cos(p_y b) ,
 \end{equation}
we have to redefine the corresponding coherence lengths in the following
way:
\begin{equation}
\tilde{\xi_x} = \xi_x , \ \ \tilde{\xi_y} = \xi_y / \sqrt{2}, \ \ \tilde{\xi_z} = \xi_z .
\end{equation}
In terms of the redefined coherence lengths the GL slopes of the anisotropic 
upper critical field for d-wave like superconducting order parameter (17) can
be expressed as
\begin{equation}
H^a_{c2} = \frac{\phi_0}{2 \pi \tilde{\xi_y} \tilde{\xi_z}} \biggl( \frac{T_c-T}{T_c} \biggl) =
\frac{4 \sqrt{2} \pi^2 c \hbar T^2_c}{7 \zeta(3) e t_b t_c b c^*} \biggl( \frac{T_c-T}{T_c}
\biggl) ,
\end{equation}
\begin{equation}
H^b_{c2} = \frac{\phi_0}{2 \pi \tilde{\xi_x} \tilde{\xi_z}} \biggl( \frac{T_c-T}{T_c} \biggl) =
\frac{8 \pi^2 c \hbar T^2_c}{7 \zeta(3) e t_a t_c a c^*} \biggl( \frac{T_c-T}{T_c}
\biggl) ,
\end{equation}
\begin{equation}
H^c_{c2} = \frac{\phi_0}{2 \pi \tilde{\xi_x} \tilde{\xi_y}} \biggl( \frac{T_c-T}{T_c} \biggl) =
\frac{8 \sqrt{2} \pi^2 c \hbar T^2_c}{7 \zeta(3) e t_a t_b a b} \biggl( \frac{T_c-T}{T_c}
\biggl) .
\end{equation}

It is important that the GL slopes of the upper critical magnetic fields along
${\bf b}$ and ${\bf c^*}$ axes have been recently carefully experimentally 
measured in the superconductor 
(TMTSF)$_2$ClO$_4$ $^{28,29}$. 
As to the GL slope for ${\bf H} \parallel {\bf a}$, it is still experimentally 
ill defined.
The latter fact is due to rather strong paramagnetic destructive effect against superconductivity, which do not allow to define carefully the orbital upper critical 
field along
${\bf a}$ axis.
Therefore, to determine the band parameters of Q1D electron spectrum (1), we need 
one more piece of information.
It is provided by theoretical fitting$^{15}$ of the LNL angular oscillations in a metallic phase of the (TMTSF)$_2$ClO$_4$ in a magnetic field.
As a result, we use the following set of experimental data$^{28,29,15}$,
\begin{equation}
\biggl( \frac{d H^b_{c2}}{d T} \biggl)_{T_c} = 3.65 \ \frac{T}{K} , 
\  \biggl( \frac{d H^c_{c2}}{d T} \biggl)_{T_c} = 0.138 \ \frac{T}{K}, 
\ t_a/t_b = 10 ,
\end{equation}
to determine all 3 band parameters in Q1D electron spectrum (1).

\begin{table}
\caption{\label{tab:table4} Band parameters of Q1D electron spectrum (1)
and critical magnetic field for the quantum $3D \rightarrow 2D$ dimensional 
crossover, $H^*$, determined for $t_a/t_b = 10$ $^{15}$ by means of Eqs.(5),(6),(12),(14),(20),(21).
}
\begin{ruledtabular}
\begin{tabular}{ccddd}
Superconductivity type&$t_a(K)$&t_b(K)&t_c(K)&H^*(T)\\
\hline
d-wave nodal
  &1340&134&2.0&3-6\\
 d-wave nodeless
  &1127&112.7&2.34&4-8\\
\end{tabular}
\end{ruledtabular}
\end{table}

The results of our calculations by means of Eqs.(12),(14),(20),(21) are
summarized in Table 1, where we consider two scenarios of superconductivity
in (TMTSF)$_2$ClO$_4$: d-wave nodal$^{21,25,30-32}$ and d-wave 
nodeless$^{33,34}$ ones.
Although we think that the d-wave nodal scenario is much more probable
one$^{30}$, we present also the results of our calculations for d-wave
nodeless  scenario, since we cannot completely exclude it at this 
point. [We note that the d-wave nodeless scenario is mathematically equivalent
to the considered above s-wave one.] 
In Table 1, we also present calculations of the critical magnetic field,
corresponding to the quantum $3D \rightarrow 2D$ dimensional crossover 
by means of Eqs.(5),(6).
As it follows from Table 1, the quantum dimensional crossover happens at 
magnetic fields, $H \geq H^* \simeq 3-8 \ T$, which are much lower than that
previously accepted.

Let us discuss possible experimental consequences of low value of the
critical field, responsible for $3D \rightarrow 2D$ dimensional crossover, 
$H^* \simeq 3-8 \ T$.
In this case, as shown in Refs.21,34, superconductivity can survive in a
form the hidden Reentrant superconducting phase in a magnetic field,
which is higher than both the quasi-classical upper critical field$^{35,36}$
and Clogston paramagnetic limit$^{37}$.
In particular, in (TMTSF)$_2$ClO$_4$ compound, the hidden Reentrant
superconductivity, as shown$^{21}$, can exist up to $H = 6 \ T$.
The expected quantum dimensional crossover has to change dramatically 
also metallic properties of (TMTSF)$_2$ClO$_4$ conductor  at 
$H \geq H^* \simeq 3-8 \ T$, if a magnetic field is applied parallel to its conducting 
plane and perpendicular to its conducting chains. 
Note that there already exist some preliminary experimental data in favor
of this conclusion.
Indeed, in Ref. 38, magnetoresistance of (TMTSF)$_2$ClO$_4$ conductor 
is studied in the above mentioned geometry.
In particular, it is found that, at $H \geq 3 \ T$, the magnetoresistance does
not follow the expected in quasi-classical theory$^{39}$ 
$H^2$-dependence.
There exist also another evidence of importance of the quantum 
$3D \rightarrow 2D$ dimensional crossover for metallic properties of
(TMTSF)$_2$ClO$_4$.
It is a failure of the quasi-classical theory$^{39}$ to explain the LMA minimum, experimentally observed at ${\bf H} \parallel {\bf b}$ (see, for example,
Fig.2 in Ref. 39).

As it follows from the above discussion, it is important to create a quantum 
theory of magnetoresistance in a metallic phase under the quantum
$3D \rightarrow 2D$ dimensional crossover condition (5).
We anticipate that this theory will be very challenging and cannot be obtained 
by generalizing of the existing methods.
We also pay attention that (TMTSF)$_2$ClO$_4$ conductor is very clean,
where an inverse impurity scattering time is estimated as 
$\hbar / \tau \sim 0.1 \ K$ (see Ref. 26) and, thus, 
$\hbar / \tau \ll t_c \simeq 2-2.5 \ K$.
Therefore, in this case, for estimation of a magnetic field, corresponding to 
$3D \rightarrow 2D$ dimensional crossover (4),(5),(6), we can use the physical 
picture of a coherent electron motion between the conducting planes, in contrast 
to the so-called weak-coherent regime$^{40}$.

We are thankful to N.N. Bagmet for useful discussions.
This work was supported by the NSF under Grants Nos. DMR-0705986 and 
DMR-1104512.


\begin{references}
\bibitem{Lebed-80}
As shown in Ref.[15], the so-called anion gap [3,4] in electron
spectrum of (TMTSF)$_2$ClO$_4$ material is  much less than the 
parameter $t_b$.
Therefore, we disregard the anion gap in the electron spectrum (1).
\end{references}
\end{document}